\documentclass[fleqn,twoside]{article}
\usepackage{espcrc2}

\usepackage{graphicx}
\usepackage[figuresright]{rotating}

\newcommand{\AmS}{{\protect\the\textfont2
  A\kern-.1667em\lower.5ex\hbox{M}\kern-.125emS}}

\title{NSI can improve LMA predictions:
neutrino decay in solar matter?}

\author{Jo\~{a}o Pulido\address[cftp]{Centro de F\'{\i}sica Te\'{o}rica de Part\'{\i}culas\\
Instituto Superior T\'{e}cnico Avenida Rovisco Pais, 1\\ 
1049-001 Lisbon Portugal}\thanks{pulido@cftp.ist.utl.pt},
C.R. Das\addressmark[cftp]\thanks{crdas@cftp.ist.utl.pt,
C. R. Das gratefully acknowledges a scholarship from Funda\c{c}\~{a}o para
a Ci\^{e}ncia e Tecnologia ref. SFRH/BPD/41091/2007.
This work was partially supported by Funda\c{c}\~{a}o para a  
Ci\^{e}ncia e a
Tecnologia (FCT, Portugal) through the projects CERN/FP/109305/2009,
PTDC/FIS/098188/2008 and CFTP-FCT Unit 777 which are partially funded
through POCTI (FEDER).}}
       
\begin{document}

\begin{abstract}
We investigate the prospects for non-standard interactions (NSI) in solar neutrino propagation 
and detection and find that these may solve the tension between the observed flatness of the 
SuperKamiokande electron spectrum and its large mixing angle (LMA) prediction which has a clear 
negative slope. Also the Cl rate prediction from NSI comes within 1$\sigma$ of its experimental 
value instead of the 
LMA one which lies more than 2$\sigma$ above. A remarkable consequence of NSI for solar neutrinos
is the possibility of neutrino decay into majorons and antineutrinos whose appearance probability 
is calculated but found to be rather small.
\vspace{1pc}
\end{abstract}

\maketitle


Neutrino non-standard interactions (NSI) have been introduced long ago 
\cite{Guzzo:1991hi,Roulet:1991sm} to account for a possible alternative solution to the solar 
neutrino problem. Since then a great deal of effort has been dedicated to study its possible
consequences \cite{Berezhiani:2001rt,Bolanos:2008km}.

On the other hand, although apparently dormant in the past few years, solar neutrinos
is by no means a closed subject. In fact its low energy sector is still poorly known and
the large mixing angle (LMA) predicted Cl rate lies in the range $(2.9-3.1)$ solar neutrino 
units (SNU) in comparison 
with $2.56\pm0.21$ SNU. Moreover and more importantly, there is a clear discrepancy between 
the LMA predicted spectrum for SuperKamiokande \cite{:2008zn} and the data (see fig.1).
\begin{figure}[htb]
\includegraphics[width=18pc]{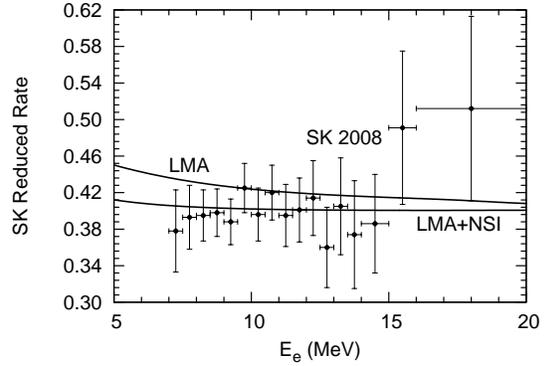}
\vspace{-1.5cm}
\caption{ \it Predictions for SuperKamiokande. 
The upper and lower curves are the LMA spectra without and with NSI. 
They are superimposed on the data published by the Collaboration in 2008 \cite{:2008zn}.}
\label{fig:toosmall}
\end{figure}
NSI are introduced as extra contributions to the vertices $\nu_{\alpha},\nu_{\beta}$ and $\nu_{\alpha} e$
so that
\begin{equation}
{\cal L}^{M}_{NSI}=-2\sqrt{2}G_F \varepsilon^{fP}_{\alpha \beta}[\bar f\gamma^{\mu}
P f][\bar\nu_{\alpha}\gamma_{\mu}P_L \nu_{\beta}]
\end{equation}
where the NSI parameters $\varepsilon^{fP}$ quantify the deviation from the standard model:
$\varepsilon\simeq \Lambda^2_{EW}/\Lambda^2_{NP}\simeq 10^{-2}$ for d=6 or
$\varepsilon\simeq \Lambda^4_{EW}/\Lambda^4_{NP}\simeq 10^{-4}$ for d=8 operators
respectively.

Treating NSI vertices like the standard interactions we get for the 
NSI potential
\begin{eqnarray}
V(NSI)=G_F\sqrt{2}~N_e\left[\left(\varepsilon_{\alpha \beta}^{eP}\right)_{CC}+\right.\nonumber\\
\left(-~\frac{1}{2}+2~sin^2\theta_W\right)\left(\varepsilon_{\alpha \beta}^{eP}\right)_{NC}+\nonumber\\
\left(1-\frac{8}{3}sin^2\theta_W+\frac{N_n}{2~N_e}\right)\varepsilon_{\alpha \beta}^{uP}+\nonumber\\
\left.\left(-\frac{1}{2}+\frac{2}{3}~sin^2\theta_W-\frac{N_n}{N_e}\right)
\varepsilon_{\alpha \beta}^{dP} \right]
\end{eqnarray}
where $N_e$ and $N_n$ are respectively the electron and neutron solar densities.
The standard inte-raction (SI) potential is
on the other hand
\begin{equation}
V(SI)=G_F\sqrt{2}N_e\left(1-\frac{N_n}{2N_e}\right)
\end{equation}
and the full one is the sum (2) and (3). Notice that $\lim_{\varepsilon_{\alpha\beta}\rightarrow 0} 
[V(SI)+V(NSI)]=V(SI)$.

Denoting by $v_{\alpha\beta}$ ($\alpha,\beta=\nu_e,\nu_{\mu},\nu_{\tau}$) the matrix elements (2), the
matter Hamiltonian in the weak basis is therefore
\begin{eqnarray}
{\cal H}_M&=&V_c \left(\begin{array}{ccc} 1 & 0 & 0 \\
0 & 0 & 0 \\
0 & 0 & 0 \\ \end{array}\right)\nonumber\\
&&+\left(\begin{array}{ccc} v_{ee} & v_{e\mu} & v_{e\tau} \\
v_{\mu e} & v_{\mu\mu} & v_{\mu\tau} \\
v_{\tau e} & v_{\tau\mu} & v_{\tau\tau} \\ \end{array}\right)^{(NSI)}
\end{eqnarray}
and in the mass basis, including the free part,
\begin{equation}
{\cal H}=\left(\begin{array}{ccc} 0 & 0 & 0 \\
0 & \frac{\Delta m^2_{21}}{2E} &  0 \\
0 &       0         &  \frac{\Delta m^2_{31}}{2E} \\ \end{array} \right)+ U^{\dagger}~
{\cal H}_M~ U~
\end{equation}
where $U$ is the Pontecorvo Maki Nakagawa Sakata (PMNS) matrix. The survival and conversion 
probabilities $(P_{ee},P_{e\mu},P_{e,\tau})$
are obtained from the integration of the Schr\"{o}dinger like equation with $\cal{H}$ and  
for the neutrino detection with electron scattering, as in SuperKamiokande, we 
allow for the possibility of $\nu_{\alpha}e^{-}\rightarrow \nu_{\beta}e^{-}$. The differential 
cross section for the process is the standard one with the replacements for the weak couplings given 
in ref. \cite{Das:2010sd} together with the spectral event rate.

Investigating the parameter range $|\varepsilon_{\alpha\beta}|~\epsilon~[5\times 10^{-5}~,~5\times 10^{-2}]$,
our analysis shows that only imaginary diagonal couplings change the LMA survival probability in the 
desired way: a flat shape for neutrino energy above $4~MeV$ while keeping the same values for low energies.
An infinite number of combinations for the $\varepsilon$ parameters is of course possible for the best fit
to the data \cite{Das:2010sd} and we choose here a common value $\varepsilon=3.5\times 10^{-4}$.

A necessary consequence of these facts is neutrino decay in solar matter. Since radiative
decay and neutrino spin-light \cite{Lobanov:2004um} are excluded in the sun, we are left with the possibility 
of decay into an antineutrino and a majoron. This decay probability as a function of antineutrino energy 
$E_f$ is given by \cite{Das:2010sd}
\begin{equation}
P_{\bar\nu_{\beta}}(E_f)\!=\!\frac{|g_{\alpha\beta}|^2}{8\pi}\!\!\int_{R_i}^{R_S}\!\!\!\!|V_{\alpha}(r)+V_{\beta}(r)|
I(r,E_f)dr
\end{equation}
with 
$$I(r,E_f)\!=\!\!\!\int_{E_{f}}^{E_{i_{max}}}\!\!\!\!\phi(E_i)(1-e^{-\Gamma(r,E_i)r})\frac{E_i-E_f}{{E_i}^2}dE_i.$$
Here $g_{\alpha\beta}$ are the neutrino majoron couplings, $\phi(E_i)$ is the incoming solar neutrino flux 
for energy $E_i$, $V_{\alpha,\beta}(r)$ are the NSI interaction potentials $R_i$ is the neutrino production point 
and $\Gamma(r,E_i)$ is the matter dependent decay rate. We take $E_{i_{max}}=16.5~MeV$ from the standard solar model. 
Upper limits for $g_{\alpha\beta}$ are given in the literature \cite{Lessa:2007up}.

The antineutrino probability (fig.2) is seen to be extremely small but grows rapidly as $E_f$ approaches
its lower limit. It decreases fast to zero as $E_f\rightarrow E_{i_{max}}$, since fewer neutrinos contribute in the 
upper energy range. Comparison with the Borexino \cite{Bellini:2010gn} and the KamLAND \cite{Eguchi:2003gg} upper 
bounds on solar $\bar\nu_e$ shows that our predictions lie within 6-8 orders of magnitude within the data 
\cite{Das:2010sd}. Hence any increase in experimental sensitivity will be unable to reveal a possible solar 
antineutrino flux produced from NSI.

Finally the full physical process in our model for neutrino propagation and
decay through NSI in the sun is represented in fig.3. Equation (6) we used for the Hamiltonian
does not take into account the extra physics involved in the majoron coupling and is
therefore a truncated Hamiltonian, whose hermiticity is restored once the detailed majoron
emission process is taken into account.

\begin{figure}[h]
\includegraphics[width=17.5pc]{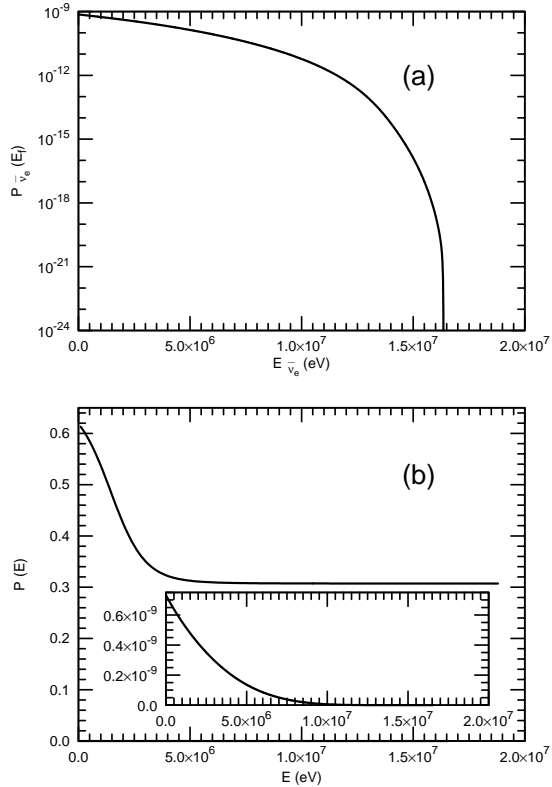}
\vspace{-1.0cm}
\caption{\it Panel (a): antineutrino production probability in a logarithmic scale. Panel (b):
LMA+NSI probability and antineutrino production probability (inner graph). Its extremely small
value prevents the possibility of observing antineutrinos from the sun from NSI.}
\label{fig:toosmall}
\end{figure}

\begin{figure}[htb]
\includegraphics[width=17pc]{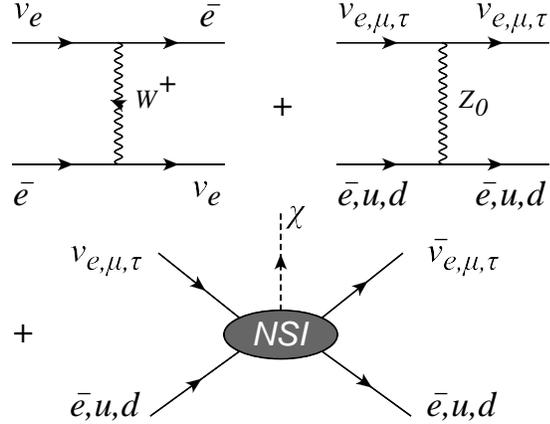}
\vspace{-0.5cm}
\caption{\it The processes involved in the propagation of neutrinos in the sun as
in our model: the two upper diagrams are the standard ones for matter oscillation and
the lower one represents the decay $\nu_i\rightarrow \bar\nu_{j}+majoron (\chi)$.}
\label{fig:toosmall}
\end{figure}

\end{document}